\documentclass[conference]{IEEEtran}

\usepackage{amsmath}
\usepackage{amssymb}
\usepackage{blindtext}
\usepackage{booktabs}
\usepackage{caption}
\usepackage{graphicx}
\usepackage[utf8]{inputenc}
\usepackage{multirow}
\usepackage[group-separator={,}]{siunitx}
\usepackage[x11names, rgb, dvipsnames]{xcolor}
\usepackage{xfrac}
\usepackage[perpage]{footmisc}
\usepackage[hidelinks=true]{hyperref}
\usepackage{microtype}
\usepackage[capitalize, nameinlink, noabbrev]{cleveref}
\usepackage{subcaption}

\usepackage{tikz}
\usetikzlibrary{shapes}
\usetikzlibrary{arrows}

\DeclareMathOperator{\ego}{Ring}
\DeclareMathOperator{\cat}{Cat}

\newcommand{\NA}{---}

\newcommand{\noimage}{%
  \setlength{\fboxsep}{-\fboxrule}
  \fbox{\phantom{\rule{\columnwidth}{100pt}}}
}
\newcommand{\includegraphicsmaybe}[1]{\IfFileExists{#1}{\includegraphics[width=\columnwidth, height=.24\textheight, keepaspectratio]{#1}}{\noimage}}

\title{Comparison of Feature Extraction Methods and Predictors for Income Inference}

\author{%
\IEEEauthorblockN{%
	Martin Fixman\IEEEauthorrefmark{1},
	Martin Minnoni\IEEEauthorrefmark{2},
	Carlos Sarraute\IEEEauthorrefmark{2}
}
\IEEEauthorblockA{\IEEEauthorrefmark{1}Universidad de Buenos Aires, Argentina}
\IEEEauthorblockA{\IEEEauthorrefmark{2}Grandata Labs, 550 15th Street, San Francisco, CA, USA}
\IEEEauthorblockA{Email: martinfixman@gmail.com, \{martin, charles\}@grandata.com}
}

\begin{document}
\maketitle

\begin{abstract}

Patterns of mobile phone communications, coupled with the information of the social network graph and financial behavior, allow us to make inferences of users' socio-economic attributes such as their income level. We present here several methods to extract features from mobile phone usage (calls and messages), and compare different combinations of supervised machine learning techniques and sets of features used as input for the inference of users' income. Our experimental results show that the Bayesian method based on the communication graph outperforms standard machine learning algorithms using node-based features.

\end{abstract}

% !TEX root = FeatureExtraction.tex

\section{Introduction}

Mobile phone datasets present a rich view into social interactions and physical movements of large segments of the population.
The voice calls and text messages exchanged between people, together with the locations of these communications, allow us to construct a rich social graph which provides insights on the users social fabric.

There is a strong homophily in the population's communications graph respect to economic variables such as the user's income~\cite{fixman2016bayesian}, which results largely from social stratification between populations of different purchasing power~\cite{leo2015socioeconomic} or purchasing patterns~\cite{Leo2016correlations}.
Additionally, and in part resulting from this stratification, there are different patterns of communication between users of distinct socioeconomic level~\cite{Luo2017inferring}.

Finding the best way to parse Call Detail Records (CDRs) to generate user features and construct their communications graph is still a subject of research.  After describing our data sources, we present several methods of feature extraction from the raw CDR data,
and describe the supervised machine learning algorithms used to predict the \emph{socioeconomic level} of users, given a ground truth for a relatively small subset of the population. In particular we tested the Bayesian approach for income inference presented in~\cite{fixman2016bayesian}.
Finally, we present our experimental results, comparing the performance obtained according to the feature set and the algorithms used.

\begin{table*}
\begin{tabular*}{\textwidth}{>{\bfseries}l >{\bfseries}l @{\extracolsep{\fill}}>{\hspace{2em}}r r r r r r >{\hspace{2em}}r >{\hspace{-1em}}r}
\multicolumn{10}{>{\bfseries}c}{Inner Graph} \\
\toprule
Model & Level & Accuracy & Precision & Recall & AUC & F\textsubscript{1}-score & F\textsubscript{4}-score & Fit Time & Predict Time \\
\midrule

\multicolumn{2}{>{\bfseries}l}{Random Selection}
& 0.499 & 0.499 & 0.500 & 0.499 & 0.500 & 0.500 & \NA{} & \SI{0.005}{\second} \\

\multicolumn{2}{>{\bfseries}l}{Majority Voting}
& 0.681 & 0.640 & 0.826 & 0.681 & 0.721 & 0.712 & \NA{} & \SI{0.059}{\second} \\

\multicolumn{2}{>{\bfseries}l}{Bayesian Algorithm}
& 0.693 & 0.665 & 0.792 & \textbf{0.746} & \textbf{0.723} & \textbf{0.783} & \NA{} & \SI{33.155}{\second} \\
\midrule

\multirow{5}{*}{LR} &
$\ego_1$ & 0.536 & 0.531 & 0.625 & 0.536 & 0.574 & 0.619 & \SI{0.145}{\second}   & \SI{0.002}{\second} \\
& $\ego_2$ & 0.535 & 0.525 & 0.730 & 0.535 & 0.611 & 0.714 & \SI{0.141}{\second}   & \SI{0.011}{\second} \\
& $\ego_3$ & 0.568 & 0.578 & 0.525 & 0.569 & 0.550 & 0.528 & \SI{0.119}{\second}   & \SI{0.003}{\second} \\
& $\cat_1$ & 0.686 & 0.655 & 0.785 & 0.686 & 0.714 & 0.776 & \SI{0.167}{\second}   & \SI{0.005}{\second} \\
& $\cat_2$ & 0.693 & 0.665 & 0.780 & 0.693 & 0.718 & 0.772 & \SI{1.588}{\second}   & \SI{0.011}{\second} \\
& $\cat_3$ & 0.693 & 0.670 & 0.764 & 0.692 & 0.714 & 0.758 & \SI{0.956}{\second}   & \SI{0.009}{\second} \\

\midrule

\multirow{5}{*}{RF} &
$\ego_1$ & 0.548 & 0.548 & 0.550 & 0.548 & 0.549 & 0.550 & \SI{5.986}{\second}   & \SI{0.588}{\second} \\
& $\ego_2$ & 0.582 & 0.583 & 0.577 & 0.582 & 0.580 & 0.577 & \SI{56.548}{\second}  & \SI{0.483}{\second} \\
& $\ego_3$ & 0.576 & 0.577 & 0.580 & 0.576 & 0.579 & 0.580 & \SI{50.197}{\second}  & \SI{0.253}{\second} \\
& $\cat_1$ & 0.671 & 0.665 & 0.690 & 0.671 & 0.677 & 0.688 & \SI{6.346}{\second}   & \SI{0.539}{\second} \\
& $\cat_2$ & \textbf{0.714} & 0.713 & 0.716 & 0.714 & 0.714 & 0.716 & \SI{96.005}{\second}  & \SI{0.460}{\second} \\
& $\cat_3$ & 0.709 & 0.710 & 0.711 & 0.709 & 0.711 & 0.711 & \SI{81.528}{\second}  & \SI{0.242}{\second} \\
\bottomrule
\end{tabular*}
\caption{Resulting metrics of different methods used in \cref{sec:results} tested on the \emph{Inner Graph}, which contains only nodes which have at least one neighbour with socioeconomic information. \textbf{LR} corresponds to \emph{Logistic Regression} models, and \textbf{RF} to \emph{Random Forest} models with the level described in \cref{sec:accumulatedfeatures}. Bolded items represent the highest value for the metrics Accuracy, AUC, F\textsubscript{1}-score and F\textsubscript{4}-score.}
\label{tab:innergraphresults}
\end{table*}

% !TEX root = FeatureExtraction.tex

\section{Data Sources}
\label{sec:data_sources}

The data used in this study contains a set $P$ of \textit{Call Detail Records} (CDRs), composed of voice calls, and another set $S$ containing text messages, from a telecommunication company (\textit{telco}) for a period of 3 consecutive months. Using this data we create the social graph $G = \left< V, E \right>$ where each $v \in V$ is a user of the telco, and $E$ contains calls between those users. Each element $e \in E$ contains information about the \emph{Origin} and \emph{Destination} users, in addition to the amount of \emph{Calls}, the total call \emph{Time}, and the amount of \emph{SMS} exchanged.

Additionally, we have access to information about a set $B$ of bank accounts, for which we calculate the monthly income for each user $p_s$. In this paper we separate the users into two groups of equal size: \emph{Low Income} and \emph{High Income}.

$B$ contains information about the users' telephone number, which is anonymized in the same way as the telco data.  Therefore, we can match the data in these two datasets in order to construct the \emph{Ground Truth} $T \subseteq V$, where each element of $T$ contains its income category, along with the \emph{Inner Graph} $G' = \left< V', E' \right>$ where E' contains edges where at least one endpoint is in $T$, and $V'$ is the set of endpoints of all elements in $E'$.

% !TEX root = FeatureExtraction.tex

\section{Accumulated Graph Features}
\label{sec:accumulatedfeatures}

This section presents several ways of transforming data from the graph $G = \left< V, E \right>$ into individual features for each user $v \in V$.
The aggregations are classified into levels named according to the transformation done to $G$, and they are merged with levels containing less information.
The total amount of columns in each featureset is presented in \cref{tab:features}.

\begin{table}[t]
\centering
\begin{tabular}{>{\bfseries}l r}
\toprule
Level & Features \\
\midrule
$\ego_1$ & \num{8}  \\
$\ego_2$ & \num{16} \\
$\ego_3$ & \num{24} \\
$\cat_1$ & \num{24} \\
$\cat_2$ & \num{48} \\
$\cat_3$ & \num{72} \\
\bottomrule
\end{tabular}
\caption{Amount of total features per level.}
\label{tab:features}
\end{table}

\subsection{User Data --- Level $\ego_1$}
\label{subsec:user_data}

The first accumulated features consist of accumulating the three quantifiable features, \emph{Calls}, \emph{Time} and \emph{SMS}, for every node, separated on whether those features are incoming or outgoing.

\subsection{Higher Order User Data --- Level $\ego_{n > 1}$}
\label{subsec:higherorderuserdata}

The \emph{ego network} of the node $v$ is defined as the graph consisting of $v$ and its neighbors. A simple way to get more features about that node is to accumulate the calls and SMS information about the edges which are \textbf{not} part of the ego network, but contain one endpoint on the border of the ego network.

Similarly, we define the \emph{user data of order $n$}, for any natural number $n$, as the accumulation of calls and SMS information for the nodes which are part of the \emph{ego network of order $n$}, which is the set of nodes which are at distance at most $n$ of $v$, and are not part of the \emph{ego network of order $n - 1$}, which we denote as $\ego_n$.
The level $\ego_1$ contains the information of the regular \emph{user data}, % from \cref{subsec:user_data}, 
while the user data from the \emph{ego network of order $n$} is assigned to $\ego_n$ for $n > 1$.

\subsection{Categorical User Data --- Level $\cat_n$}
\label{subsec:categoricaluserdata}

Another approach to building features is to do an approach similar to the \emph{user data} presented in \cref{subsec:user_data}, but further discriminating each feature which corresponds to each node $v \in V$ and each edge $e \in E$:
when $t \in T$ is the other endpoint of the link $e$, we discriminate whether $t$ corresponds to a user with high or low income.
The resulting new features are of the form represented by the set in \cref{eq:matcatuserdata}. This way we create the datasets $\cat_1 \dots \cat_3$ by using the growing ego networks $\ego_1 \dots \ego_3$.

\begin{equation}
\begin{Bmatrix} in \\ out \end{Bmatrix}
\times
\begin{Bmatrix} calls \\ time \\ sms \\ contacts \end{Bmatrix}
\times
\begin{Bmatrix} low \\ high \end{Bmatrix}
\label{eq:matcatuserdata}
\end{equation}

To prevent overfitting, the set $T$ is partitioned into two disjoint sets, $G$ and $H$, where $G$ contains roughly 75\% of the nodes in $T$ is used to calculate the features, while $H$ contains the other 25\% and is used to train the models.

% !TEX root = FeatureExtraction.tex

\section{Inference Methodology}
\label{sec:inference_methodology}

The \emph{Inner Graph} is defined so that a node $h \in H$ is part of it if and only if there is an edge $\left< h, x \right> \in E$ or $\left< x, h \right> \in E$ such that $x \in H$.
This later definition becomes important when doing inferences on features using the \emph{Categorical User Data} dataset.

The inferences were performed using \emph{Logistic Regression} and \emph{Random Forest} classifiers, both of which are solid classifiers commonly used for cases like this~\cite{binaryevaluation}, and since they tend to have different variance in their results~\cite{ting2016}, noise from different sources should affect differently each predictor.

The features used were the ones presented in \cref{sec:accumulatedfeatures}, where each level is merged with all the previous levels with the data on $G$. \cref{tab:features} shows the amount of features in each level after merging the data. 
The feature sets $\ego_1 \ldots \ego_3$ and $\cat_1 \ldots \cat_3$ are strictly
increasing.

The classifiers are trained using those features and the labels in $H$ doing a \emph{Grid Search} on different hyperparameters of the predictors with \emph{5-fold cross-validation} to prevent overfitting. In addition to the \emph{accuracy}, we computed several metrics for the comparison. In our real-life use cases, we are more interested in having high \emph{recall} than high \emph{precision} (that is, finding more high income users than being accurate), therefore we also measured the \emph{F\textsubscript{4}-score} of each prediction.

In addition, these methods based on features aggregated by node are compared against three other methods based solely on the communication graph structure:

\begin{itemize}
	\item \textbf{Random Selection} which chooses a random category.
	\item \textbf{Majority Voting} which chooses the most populated category in a user's ego network (or randomly in case of a tie).
	\item \textbf{Bayesian Method} which uses the method presented in~\cite{fixman2016bayesian} to infer the category of each user,
	taking into account the uncertainty on the available information.
\end{itemize}

% !TEX root = FeatureExtraction.tex

\section{Results and Conclusion}
\label{sec:results}

\begin{table*}
\begin{tabular*}{\textwidth}{>{\bfseries}l >{\bfseries}l @{\extracolsep{\fill}}>{\hspace{2em}}r r r r r r >{\hspace{2em}}r >{\hspace{-1em}}r}
\multicolumn{10}{>{\bfseries}c}{Full Graph} \\
\toprule
Model & Level & Accuracy & Precision & Recall & AUC & F\textsubscript{1}-score & F\textsubscript{4}-score & Fit Time & Predict Time \\
\midrule

\multicolumn{2}{>{\bfseries}l}{Random Selection}
& 0.499 & 0.499 & 0.500 & 0.499 & 0.500 & 0.500 & \NA{} & \SI{0.005}{\second} \\

\multicolumn{2}{>{\bfseries}l}{Majority Voting}
& 0.565 & 0.747 & 0.197 & 0.565 & 0.312 & 0.206 & \NA{} & \SI{0.204}{\second} \\
\midrule

\multirow{5}{*}{LR}
& $\ego_1$ & 0.534 & 0.586 & 0.234 & 0.534 & 0.335 & 0.243 & \SI{0.937}{\second}   & \SI{0.016}{\second} \\
& $\ego_2$ & 0.547 & 0.617 & 0.250 & 0.547 & 0.356 & 0.260 & \SI{1.347}{\second}   & \SI{0.035}{\second} \\
& $\ego_3$ & 0.563 & 0.586 & 0.430 & 0.563 & 0.496 & 0.437 & \SI{1.055}{\second}   & \SI{0.023}{\second} \\
& $\cat_1$ & 0.565 & 0.746 & 0.198 & 0.565 & 0.313 & 0.207 & \SI{1.871}{\second}   & \SI{0.041}{\second} \\
& $\cat_2$ & 0.577 & 0.727 & 0.247 & 0.577 & 0.368 & 0.257 & \SI{9.816}{\second}   & \SI{0.077}{\second} \\
& $\cat_3$ & 0.589 & 0.636 & 0.415 & 0.589 & 0.503 & 0.424 & \SI{9.456}{\second}   & \SI{0.065}{\second} \\
\midrule

\multirow{5}{*}{RF}
& $\ego_1$ & 0.543 & 0.544 & 0.529 & 0.543 & 0.536 & 0.530 & \SI{25.789}{\second}  & \SI{4.878}{\second} \\
& $\ego_2$ & 0.578 & 0.585 & 0.537 & 0.578 & 0.560 & 0.540 & \SI{102.961}{\second} & \SI{5.608}{\second} \\
& $\ego_3$ & 0.583 & 0.590 & 0.541 & 0.583 & 0.564 & 0.543 & \SI{70.447}{\second}  & \SI{3.148}{\second} \\
& $\cat_1$ & 0.568 & 0.573 & 0.536 & 0.568 & 0.554 & 0.538 & \SI{32.981}{\second}  & \SI{5.371}{\second} \\
& $\cat_2$ & 0.613 & 0.634 & 0.533 & 0.613 & 0.579 & 0.538 & \SI{44.911}{\second}  & \SI{6.002}{\second} \\
& $\cat_3$ & \textbf{0.614} & 0.635 & 0.534 & \textbf{0.614} & \textbf{0.580} & \textbf{0.539} & \SI{50.589}{\second}  & \SI{3.484}{\second} \\
\bottomrule
\end{tabular*}
\caption{Resulting metrics of different methods used in \cref{sec:results} tested on both the \emph{Full Graph}, which includes all the nodes of the graph. \textbf{LR} corresponds to \emph{Logistic Regression} models, and \textbf{RF} to \emph{Random Forest} ones with the level described in \cref{sec:accumulatedfeatures}. Bolded items represent the highest value for the metrics Accuracy, AUC, F\textsubscript{1}-score and F\textsubscript{4}-score.}
\label{tab:fullgraphresults}
\end{table*}

The predictors were run in a computer with a single core of 2.00 GHz Intel Xeon CPU using \texttt{sklearn 0.18} under \texttt{Python 2.7}, and 64 GB of RAM (enough to avoid caching calculations). The results of the inference can be found in \cref{tab:innergraphresults} for the \emph{Inner Graph} and in \cref{tab:fullgraphresults} for the \emph{Full Graph}.

Both tables show various metrics which result from applying the methods described in \cref{sec:inference_methodology} with the hyperparameters that result in the highest \emph{accuracy} according to the \emph{Grid Search}.
We use the \emph{AUC} (Area under the ROC curve) to compare the different approaches.

We observe that methods based on \emph{Random Forests} tend to perform better in this real-world scenario than the ones based on \emph{Logistic Regression}.
This may be due to the fact that \emph{Random Forests} are more versatile with non-linear data~\cite{logisticvsdecision}, and is consistent with similar findings in~\cite{muchlinski2016}.

Interestingly, increasing the breadth of the \emph{Ego Network} by one level, from $\ego_1$ to $\ego_2$ improves the performance when using \emph{Random Forest} learning, however it does not improve by going one level further to $\ego_3$ in the case of the \emph{Inner Graph}, despite the fact that this dataset is a strict superset of the previous $\ego_2$. This is due to the fallibility of common \emph{bagging methods} like \emph{Random Forest}, where having some noisy or non-informative data to choose from makes it less probable that informative features will be chosen.

Adding categorical information greatly improves the prediction when using either method, particularly on \emph{Random Forest}, and, like it was noted before, adding neighbouring data of the \emph{ego network of distance 2} ($\cat_2$) also results in a better predictor. However, raising further the maximum distance to $\cat_3$ within the {ego network} in the case of the \emph{Inner Graph} does not result in further improvements.

We can conclude that, within the {machine learning} methods presented, the best in terms of \emph{AUC} is predicting the category using a \emph{Random Forest} with the \emph{ego network of distance 2} data ($\cat_2$) in the case of the \emph{Inner Graph}. In the \emph{Full Graph}, using $\cat_3$ data results in slightly better results, however the difference with $\cat_2$ is very small.

Finally, in the \emph{Inner Graph} the best method is the \emph{Bayesian Method} presented in~\cite{fixman2016bayesian}, which only uses the number of \emph{High Income} and \emph{Low Income} users in the egonetwork, but makes a ``smarter'' prediction than the {machine learning} methods LR and RF using the models $\cat_1$, $\cat_2$, and $\cat_3$, which also contain this data. 

Additionally, while its AUC is not higher than the best {machine learning} methods, the F\textsubscript{1}-score of the \emph{Majority Voting} predictor is higher than all the {machine learning} methods.

We can reach the conclusion that, in this particular case, \emph{smaller is better}. The {machine learning} methods which use many features (despite these features being informative) are not better at predicting the socioeconomic level of a user than the simple \emph{Majority Voting} or the more complex \emph{Bayesian Method} which use only 2 simple features of the communication graph.

\bibliography{../bibliography/sna}{}

\end{document}